# Extending mixtures of factor models using the restricted multivariate skew-normal distribution

Tsung-I Lin,* Geoffrey J. McLachlan, Sharon X. Lee

**Abstract**

The mixture of factor analyzers (MFA) model provides a powerful tool for analyzing high-dimensional data as it can reduce the number of free parameters through its factor-analytic representation of the component covariance matrices. This paper extends the MFA model to incorporate a restricted version of the multivariate skew-normal distribution to model the distribution of the latent component factors, called mixtures of skew-normal factor analyzers (MSNFA). The proposed MSNFA model allows us to relax the need for the normality assumption for the latent factors in order to accommodate skewness in the observed data. The MSNFA model thus provides an approach to model-based density estimation and clustering of high-dimensional data exhibiting asymmetric characteristics. A computationally feasible ECM algorithm is developed for computing the maximum likelihood estimates of the parameters. Model selection can be made on the basis of three commonly used information-based criteria. The potential of the proposed methodology is exemplified through applications to two real examples, and the results are compared with those obtained from fitting the MFA model.

*Key words*: clustering; data reduction; ECM algorithm; factor analyzer; rMSN distribution; skewness

*Tsung-I Lin

Institute of Statistics, National Chung Hsing University, Taichung 402, Taiwan;
e-mail: tilin@nchu.edu.tw

Geoffrey J. McLachlan · Sharon X. Lee

Department of Mathematics, University of Queensland, St Lucia, 4072, Australia

# 1 Introduction

Factor analysis (FA) is a popular technique for explaining the covariance relationships among many variables through a fewer number of unobservable random quantities known as *latent factors*. Finite mixture models (FMMs) are being widely used as a flexible means to model heterogeneous data, in particular, for density estimation and clustering. There are a number of monographs on mixture models; see, for example, Everitt and Hand (1981), Titterington et al. (1985), McLachlan and Basford (1988), Lindsay (1995), Böhning(1999), McLachlan and Peel (2000a), Frühwirth-Schnatter (2006), and Mengersen et al. (2011), and the references contained therein. Mixtures of factor analyzers (MFAs) were introduced by Ghahramani and Hinton (1997). They provide a global non-linear approach to dimension reduction via the adoption of component distribtions having a factor-analytic representation for the component-covariance matrices; see also McLachlan and Peel (2000b). McLachlan et al. (2002, 2003) exploited the MFA model for the analysis of high-dimensional data, including the clustering of microarray gene-expression profiles. For data with clusters having tails longer than the normal distribution, McLachlan et al. (2007) adopted the family of multivariate $t$-distributions for the component factors and errors to establish a robust extension of MFA. More recently, Baek et al. (2010) proposed mixtures of common factor analyzers in which the factors are takne to have a common distribution before thier transformation to be white noise. A robust version of this approach using $t$-component distributions was subsequently provided by Baek et al. (2011). Bayesian treatments of the MFA model have been investigated by Ghahramani and Beal (2000) via a variational approximation and Utsugi and Kumagai (2001) using the Gibbs sampler and a deterministic algorithm; see also Mengersen et al. (2011).

For computational convenience and mathematical tractability, component errors and latent factors in the traditional MFA model are routinely assumed to follow multivariate normal distributions. However, in many applied problems, the data to be analyzed may contain a group or groups of observations whose distributions are moderately or severely skewed. Just like other normal-based mixture models, a slight deviation from normality may seriously affect the estimates of mixture parameters and/or lead to spurious groups, subsequently misleading inference from data. Wall et al. (2012) conducted several simulation studies to explore the influence of non-normal latent factors in the estimation of parameters.

To allow for the modeling real data as appropriately as possible and to remedy unrealistic assumptions in classical normal based multivariate version was studied by Azzalini and Dalla Valle (1996), Azzalini and Capitanio (1999), Gupta et al. (2004), and Arellano-Valle and Genton (2005), among others. In recent years, there has been growing interest in the study of mixtures of skew-normal distributions (Lin et al., 2007; Lin, 2009), both in the univariate and multivariate cases, as a more general tool for handling heterogeneous data involving asymmetric behavior across sub-populations. Pyne et al. (2009) proposed another finite mixture model with multivariate skew-normal or $t$-distributions based on a restricted variant of the skew-elliptical family of distributions of Sahu et al. (2003), one of which is referred to as the restricted multivariate skew-normal (rMSN) distribution. Lee and McLachlan (2013a, 2013b) have provided a systematic overview of various existing multivariate skew distributions and clarified their conditioning-type and convolution-type representations. Also, Lee and McLachlan (2013c) have provided the package EMMIX-uskew, which implements a closed-form expectation-maximization (EM) algorithm for computing the ML estimates of the parameters for mixtures of restricted and unrestricted skew-normal and skew $t$-distributions.

In this paper, we propose mixtures of skew-normal factor analyzers (MSNFA) where the latent component factors are assumed to follow the family of rMSN distributions in an attempt to model the data precisely in the presence of skewed sub-populations. The proposed model can be viewed as a novel dimensionally reduced model-based approach. It is a generalization of the MFA model, allowing for an appropriate representation of non-normal data. Maximum likelihood (ML) estimates of the parameters in the model can be computed via the closed-form EM implementations (Dempster et al. 1977; Meng and van Dyk, 1997), and the estimated factor scores are obtained as by-products within the estimation procedure. The asymptotic covariance matrix of estimated mixture parameters is obtained by inverting an approximation to the observed information matrix as suggested by Jamshidian (1997).

The rest of the paper is organized as follows. In Sect. 2, we establish notation and provide a preliminary account of the rMSN distribution. In Sect. 3, we briefly present the formulation of the skew-normal factor analysis (SNFA) model and study some related properties. Sect. 4 extends the work to the MSNFA model and presents an EM-type algorithm for obtaining the ML estimates of model parameters. Sect. 5 describes some practical issues, including the specification of starting values, the stopping rule, model selection and two indices for performance evaluation. The proposed methodologies are illustrated through application to two well-known real examples in Sect. 6. Some concluding remarks are given in Sect. 7.

## 2 The restricted multivariate skew-normal distribution

We begin with a brief review of a restricted version of the MSN distribution and a study of some essential properties. A unification of families of MSN distributions and several variants and extensions can be found in Azzalini (2005) and Arellano-Valle and Azzalini (2006). To establish notation, let $\phi_p(\cdot; \boldsymbol{\mu}, \boldsymbol{\Sigma})$ be the probability density function (pdf) corresponding to $N_p(\boldsymbol{\mu}, \boldsymbol{\Sigma})$, a $p$-dimensional multivariate normal distribution with mean $\boldsymbol{\mu}$ and covariance matrix $\boldsymbol{\Sigma}$, and $\Phi(\cdot)$ the cumulative distribution function (cdf) of the standard normal distribution. Further, let $TN(\mu, \sigma^2; (a, b))$ denote the truncated normal distribution for $N(\mu, \sigma^2)$ lying within a truncated interval $(a, b)$.

Following Lee and McLachlan (2013b), a $p \times 1$ random vector $\boldsymbol{X}$ is said to follow a rMSN distribution with location vector $\boldsymbol{\mu}$, dispersion matrix $\boldsymbol{\Sigma}$ and skewness vector $\boldsymbol{\lambda}$, denoted by $\boldsymbol{X} \sim rSN_p(\boldsymbol{\mu}, \boldsymbol{\Sigma}, \boldsymbol{\lambda})$, if it can be represented as

$$\boldsymbol{X} = \boldsymbol{\lambda}|U_1| + \boldsymbol{U}_2, \quad U_1 \perp \boldsymbol{U}_2, \tag{1}$$

where $U_1 \sim N(0,1)$, $\boldsymbol{U}_2 \sim N_p(\boldsymbol{\mu}, \boldsymbol{\Sigma})$ and the symbol '$\perp$' indicates independence. Letting $W = |U_1|$, a 2-level hierarchical representation of (1) is

$$\begin{aligned} \boldsymbol{X} \mid (W = w) &\sim N_p(\boldsymbol{\mu} + \boldsymbol{\lambda} w, \boldsymbol{\Sigma}), \\ W &\sim TN\left(0, 1; (0, \infty)\right). \end{aligned} \tag{2}$$

For computing the moments of $W$, we use the following result.

**Proposition 1.** *Let $W \sim TN(\mu, \sigma^2; (0, \infty))$. The density of $W$ is*

$$f(w) = \frac{\phi(w; \mu, \sigma^2)}{\Phi(\mu/\sigma)} I(w > 0),$$

*where $I(\cdot)$ is an indicator function. For positive integer $k$, the moments of $W$ are given by*

$$\begin{aligned} E(W) &= \mu + \sigma \frac{\phi(\mu/\sigma)}{\Phi(\mu/\sigma)} \quad \text{for } k = 1, \\ E(W^k) &= (k-1)\sigma^2 E(W^{k-2}) + \mu E(W^{k-1}) \quad \text{for } k \geq 2. \end{aligned}$$

The pdf of $\boldsymbol{X}$, expressed as a product of a multivariate normal density and a univariate normal distribution function, is given by

$$f(\boldsymbol{x}) = 2\phi_p(\boldsymbol{x}; \boldsymbol{\mu}, \boldsymbol{\Omega})\Phi(\xi/\sigma), \tag{3}$$

where $\boldsymbol{\Omega} = \boldsymbol{\Sigma} + \boldsymbol{\lambda}\boldsymbol{\lambda}^\top$, $\xi = \boldsymbol{\lambda}^\top \boldsymbol{\Omega}^{-1}(\boldsymbol{x} - \boldsymbol{\mu})$, and $\sigma^2 = (1 + \boldsymbol{\lambda}^\top \boldsymbol{\Sigma}^{-1} \boldsymbol{\lambda})^{-1} = 1 - \boldsymbol{\lambda}^\top \boldsymbol{\Omega}^{-1} \boldsymbol{\lambda}$. The rMSN distribution falls into the class of fundamental skew-normal (FUSN) distribution (Arellano-Valle and Genton, 2005). In addition, it can be treated as a simplified version of Sahu et al. (2003) or a modification of the traditional version of Azzalini and Dalla Valle (1996) via a reparameterization. Such a version allows us to develop computational feasible EM-type algorithms for parameter estimation in SNFA and MSNFA models.

From (1), by Proposition 1 and the law of iterative expectations, the mean and covariance matrix of $\boldsymbol{X}$ are

$$E(\boldsymbol{X}) = \boldsymbol{\mu} + c\boldsymbol{\lambda} \quad \text{and} \quad \text{cov}(\boldsymbol{X}) = \boldsymbol{\Sigma} + (1-c^2)\boldsymbol{\lambda}\boldsymbol{\lambda}^\top, \tag{4}$$

where $c = \sqrt{2/\pi}$. Higher order moments can be derived from the moment generating function (mgf) given in the following proposition.

**Proposition 2.** *If $\boldsymbol{X} \sim rSN_p(\boldsymbol{\mu}, \boldsymbol{\Sigma}, \boldsymbol{\lambda})$, then the mgf of $\boldsymbol{X}$ is*

$$M_{\boldsymbol{x}}(\boldsymbol{t}) = 2\exp\left(\boldsymbol{t}^\top\boldsymbol{\mu} + \frac{1}{2}\boldsymbol{t}^\top\boldsymbol{\Omega}\boldsymbol{t}\right)\Phi(\boldsymbol{\lambda}^\top\boldsymbol{t}), \quad \boldsymbol{t} \in \mathbb{R}^p.$$

The following result shows an appealing closure property of the rMSN distribution under affine transformation, which is useful for later methodological developments.

**Proposition 3.** *Let $\boldsymbol{X} \sim rSN_p(\boldsymbol{\mu}, \boldsymbol{\Sigma}, \boldsymbol{\lambda})$. For any full rank matrix $\boldsymbol{L} \in \mathbb{R}^{q \times p}$ ($1 \leqslant q \leqslant p$), the distribution of the linear transformation $\boldsymbol{LX}$ is*

$$\boldsymbol{LX} \sim rSN_q(\boldsymbol{L\mu}, \boldsymbol{L\Sigma L}^\top, \boldsymbol{L\lambda}).$$

The proof follows easily by applying Proposition 2 to the transformation $\boldsymbol{LX}$.

## 3 The skew-normal factor analysis model

### 3.1 The model

We consider a generalization of the traditional FA model in which the hidden factors are assumed to follow an rMSN distribution within the family defined by (1). Suppose that $\boldsymbol{Y} = \{\boldsymbol{Y}_1, \ldots, \boldsymbol{Y}_n\}$ is a random sample of $n$ $p$-dimensional observations. The SNFA model considered can be written as

$$\begin{cases} \boldsymbol{Y}_j = \boldsymbol{\mu} + \boldsymbol{B}\boldsymbol{U}_j + \boldsymbol{\varepsilon}_j, & \boldsymbol{U}_j \perp \boldsymbol{\varepsilon}_j \\ \boldsymbol{U}_j \stackrel{\text{iid}}{\sim} rSN_q(-c\boldsymbol{\Delta}^{-1/2}\boldsymbol{\lambda}, \boldsymbol{\Delta}^{-1}, \boldsymbol{\Delta}^{-1/2}\boldsymbol{\lambda}), & \boldsymbol{\varepsilon}_j \stackrel{\text{iid}}{\sim} N_p(\boldsymbol{0}, \boldsymbol{D}), \end{cases} \quad (5)$$

for $j = 1, \ldots, n$, where $\boldsymbol{\mu}$ is a $p$-dimensional location vector, $\boldsymbol{B}$ is a $p \times q$ matrix of factor loadings, $\boldsymbol{U}_j$ is a $q$-dimensional vector $(q < p)$ of latent variables called *factors*, $\boldsymbol{\varepsilon}_j$ is a $p$-dimensional vector of errors and $\boldsymbol{\Delta} = \boldsymbol{I}_q + (1-c^2)\boldsymbol{\lambda}\boldsymbol{\lambda}^{\text{T}}$ is a scaling matrix. The elements of the factor loading matrix $\boldsymbol{B}$ indicate the strength of dependence of each variable on each factor. Moreover, $\boldsymbol{D}$ is a positive diagonal matrix and $\boldsymbol{I}_q$ stands for a $q$-dimensional identity matrix.

Under model (5), an appealing property is that

$$\text{E}(\boldsymbol{U}_j) = \boldsymbol{0} \quad \text{and} \quad \text{cov}(\boldsymbol{U}_j) = \boldsymbol{I}_q. \quad (6)$$

Hence, the chosen distributional assumption for $\boldsymbol{U}_j$ makes the factor score estimates of FA and SNFA models comparable. By Proposition 3, we can deduce that

$$\boldsymbol{Y}_j \sim rSN_p(\boldsymbol{\mu} - c\boldsymbol{\alpha}, \boldsymbol{\Sigma}, \boldsymbol{\alpha}),$$

where $\boldsymbol{\Sigma} = \boldsymbol{B}\boldsymbol{\Delta}^{-1}\boldsymbol{B}^{\top} + \boldsymbol{D}$ and $\boldsymbol{\alpha} = \boldsymbol{B}\boldsymbol{\Delta}^{-1/2}\boldsymbol{\lambda}$. Clearly, the marginal distribution of $\boldsymbol{Y}_j$ belongs to the family of rMSN distributions in which the skewness parameter $\boldsymbol{\alpha}$ depends both on $\boldsymbol{B}$ and $\boldsymbol{\lambda}$. It follows immediately from (4) that

$$E(\boldsymbol{Y}_j) = \boldsymbol{0} \quad \text{and} \quad \text{cov}(\boldsymbol{Y}_j) = \boldsymbol{B}\boldsymbol{B}^{\top} + \boldsymbol{D}. \quad (7)$$

Another interesting feature of this model is that the parameters estimates of $\boldsymbol{\mu}$, $\boldsymbol{B}$ and $\boldsymbol{D}$ can be used to recover the sample mean and sample covariance for both FA and SNFA models. The two important characteristics (6) and (7) were not considered by Montanrai and Viroli (2010) and in other developments in the literature.

### 3.2 Identifiability issues

For a hidden dimensionality $q > 1$, there is an identifiability issue associated with the rotation invariance of factor loading matrix $\boldsymbol{B}$. For any orthogonal matrix $P$ of order $q$, model (5) still satisfies when $B$ is replaced by $BP$ and the latent $U_j$ is changed to $P^{\mathrm{T}} U_j$. Moreover, such an orthogonal transformation will leave the covariance matrix in (7) invariant since $BP(BP)^{\mathrm{T}} = BB^{\top}$.

To circumvent this identifiability problem (rotational indeterminacy), one of the most commonly used techniques is to constrain the loading matrix $B$ so that the upper-right triangle is zero and the diagonal entries are strictly positive (e.g., Fokoué and Titterington, 2003; Lopes and West, 2004). This means that $q(q-1)/2$ elements of $B$ are constrained. We therefore have a total number of parameters $m = p(q+2) + q - q(q-1)/2$ to be estimated.

The mixture model itself poses another identifiability problem raised by relabelling of components. More precisely, the likelihood is invariant under a permutation of the class labels in parameter vectors. Therefore a label switching problem can occur when some labels of the mixture classes permute (McLachlan and Peel, 2000). However, the switching of class labels is not a concern with the general ML approach.

## 4 Mixture of restricted skew-normal factors

### 4.1 Model formulation

Let $\boldsymbol{Y}_j = (Y_{j1}, \ldots, Y_{jp})^\top$ be a $p$-dimensional vector of $p$ feature variables ($j = 1, \ldots, n$), where $\boldsymbol{Y}_j$ comes from a heterogeneous population with $g$ non-overlapping components. To denote which component $\boldsymbol{Y}_j$ belongs in this finite mixture framework, we introduce the latent membership-indicator vectors, $\boldsymbol{Z}_1, \ldots, \boldsymbol{Z}_n$. Here $Z_{ij} = (\boldsymbol{Z}_j)_i$ is zero or one, according as to whether $\boldsymbol{Y}_j$ belongs or does not belong to the $i$th component ($i = 1, \ldots, g$; $j = 1, \ldots, n$). Accordingly, we have

$$\boldsymbol{Z}_1 \ldots, \boldsymbol{Z}_n \overset{\mathrm{iid}}{\sim} \mathcal{M}(1; \pi_1, \ldots, \pi_g),$$

where the probability function of $\boldsymbol{Z}_j$ is given by

$$f(\boldsymbol{z}_j; \boldsymbol{\pi}) = \pi_1^{z_{1j}} \pi_2^{z_{2j}} \cdots \pi_g^{z_{gj}}, \quad \text{for } j = 1, \ldots, n,$$

and $\boldsymbol{\pi} = (\pi_1, \ldots, \pi_g)^\top$.

The MSNFA model is a generalization of MFA by postulating a mixture of $g$ SNFA sub-models for the distribution of the observation $\boldsymbol{Y}_j$ given the unobservable factor $\boldsymbol{U}_j$. We consider the use of MSNFA in an attempt to make the model accommodate heavy skewness arising frequently in high-dimensional data while without performing transformation.

Given $Z_{ij} = 1$, each $\boldsymbol{Y}_j$ can be modelled as

$$\boldsymbol{Y}_j = \boldsymbol{\mu}_i + \boldsymbol{B}_i \boldsymbol{U}_{ij} + \boldsymbol{\varepsilon}_{ij}, \quad \text{with probability} \quad \pi_i \quad (i = 1, \ldots, g), \tag{8}$$

for $j = 1, \ldots, n$, where the factors $\boldsymbol{U}_{i1}, \ldots, \boldsymbol{U}_{in}$ are distributed independently $rSN_q(-c\boldsymbol{\Delta}_i^{-1/2}\boldsymbol{\lambda}_i, \boldsymbol{\Delta}_i^{-1}, \boldsymbol{\Delta}_i^{-1/2}\boldsymbol{\lambda}_i)$, independently of the $\boldsymbol{\varepsilon}_{ij}$, which are distributed independently $N_p(0, \boldsymbol{D}_i)$, where $\boldsymbol{\Delta}_i = \boldsymbol{I}_q + (1-c^2)\boldsymbol{\lambda}_i\boldsymbol{\lambda}_i^\mathrm{T}$ and $\boldsymbol{D}_i$ is a positive diagonal matrix.

From (8), the marginal pdf of $\boldsymbol{Y}_j$ is

$$f(\boldsymbol{y}_j; \boldsymbol{\Theta}) = \sum_{i=1}^{g} \pi_i \psi(\boldsymbol{y}_j; \boldsymbol{\theta}_i),$$

where $\psi(\boldsymbol{y}_j; \boldsymbol{\theta}_i)$ is the pdf of rMSN distribution defined in (3), $\boldsymbol{\theta}_i = (\boldsymbol{\mu}_i, \boldsymbol{B}_i, \boldsymbol{\lambda}_i, \boldsymbol{D}_i)$ is composed of the unknown parameters of the $i$th mixture component and $\boldsymbol{\Theta} = (\pi_1, \ldots, \pi_{g-1}, \boldsymbol{\theta}_1, \ldots, \boldsymbol{\theta}_g)$ represents the entire unknown parameters. Given a set of $n$ observations $\boldsymbol{y} = \{\boldsymbol{y}_1, \ldots, \boldsymbol{y}_n\}$, ML estimation can be undertaken for this model by maximizing the log likelihood function for $\boldsymbol{\Theta}$,

$$\ell(\boldsymbol{\Theta}; \boldsymbol{y}) = \sum_{j=1}^{n} \log \left( \sum_{i=1}^{g} \pi_i \psi(\boldsymbol{y}_j; \boldsymbol{\theta}_i) \right). \tag{9}$$

Unfortunately, it is not straightforward to derive explicit analytical solutions for ML estimator of $\boldsymbol{\Theta}$. To cope with this obstacle, one usually resorts to the EM-type algorithm (Dempster et al., 1977), which is a popular iterative device for ML estimation in models involving latent variables or missing data.

Under model (8), it can be shown that

$$\boldsymbol{Y}_j \mid (Z_{ij} = 1) \sim rSN_p(\boldsymbol{\mu}_i - c\boldsymbol{\alpha}_i, \boldsymbol{\Sigma}_i, \boldsymbol{\alpha}_i), \tag{10}$$

where $\boldsymbol{\Sigma}_i = \boldsymbol{B}_i \boldsymbol{\Delta}_i^{-1} \boldsymbol{B}_i^\top + \boldsymbol{D}_i$ and $\boldsymbol{\alpha}_i = \boldsymbol{B}_i \boldsymbol{\Delta}_i^{-1/2} \boldsymbol{\lambda}_i$. To facilitate the derivation of our inference procedure, we adopt the following scaling transformation:

$$\tilde{\boldsymbol{B}}_i \triangleq \boldsymbol{B}_i \boldsymbol{\Delta}_i^{-1/2} \quad \text{and} \quad \tilde{\boldsymbol{U}}_{ij} \triangleq \boldsymbol{\Delta}_i^{1/2} \boldsymbol{U}_j.$$

Based on (2) and (10), a four-level hierarchical representation of model (8) is

$$\begin{aligned}
\boldsymbol{Y}_j \mid (\tilde{\boldsymbol{u}}_{ij}, w_j, Z_{ij} = 1) &\sim N_p(\boldsymbol{\mu}_i + \tilde{\boldsymbol{B}}_i \tilde{\boldsymbol{u}}_{ij}, \boldsymbol{D}_i), \\
\tilde{\boldsymbol{U}}_{ij} \mid (w_j, Z_{ij} = 1) &\sim N_q\big((w_j - c)\boldsymbol{\lambda}_i, \boldsymbol{I}_q\big), \\
W_j \mid (Z_{ij} = 1) &\sim TN\big(0, 1; (0, \infty)\big), \\
\boldsymbol{Z}_j &\sim \mathcal{M}(1; \pi_1, \ldots, \pi_g).
\end{aligned} \tag{11}$$

In the EM framework, the augmented quadruples $\{\boldsymbol{Y}_j, \boldsymbol{Z}_j, \tilde{\boldsymbol{U}}_{ij}, w_j\}_{j=1}^n$ are referred to as the complete data. By Bayes' Theorem, it suffices to show that

$$\begin{aligned}
\tilde{\boldsymbol{U}}_{ij} \mid (Z_{ij} = 1, w_j, \boldsymbol{y}_j) &\sim N_q(\boldsymbol{q}_{ij}, \boldsymbol{C}_i), \\
W_j \mid (Z_{ij} = 1, \boldsymbol{y}_j) &\sim TN\big(a_{ij}, 1 - \boldsymbol{\alpha}_i^\top \boldsymbol{\Omega}_i^{-1} \boldsymbol{\alpha}_i; (0, \infty)\big),
\end{aligned} \tag{12}$$

where $\boldsymbol{q}_{ij} = \boldsymbol{C}_i \big[\boldsymbol{v}_{ij} + \boldsymbol{\lambda}_i(w_j - c)\big]$, $\boldsymbol{v}_{ij} = \tilde{\boldsymbol{B}}_i^\top \boldsymbol{D}_i^{-1}(\boldsymbol{y}_j - \boldsymbol{\mu}_i)$, $\boldsymbol{C}_i = (\boldsymbol{I}_q + \tilde{\boldsymbol{B}}_i^\top \boldsymbol{D}_i^{-1} \tilde{\boldsymbol{B}}_i)^{-1}$, $a_{ij} = \boldsymbol{\alpha}_i^\top \boldsymbol{\Omega}_i^{-1}(\boldsymbol{y}_j - \boldsymbol{\mu}_i + c\boldsymbol{\alpha}_i)$ and $\boldsymbol{\Omega}_i = \boldsymbol{\Sigma}_i + \boldsymbol{\alpha}_i \boldsymbol{\alpha}_i^\top$. As an immediate consequence, we establish the following proposition, which is crucial for the calculation of some conditional expectations involved in the proposed ECM algorithm.

**Proposition 4.** *Given the hierarchical representation (12), we have the following (the symbol "$\mid \cdots$" denotes conditioning on $Z_{ij} = 1$ and $\boldsymbol{Y}_j = \boldsymbol{y}_j$):*

(a) *The conditional expectation of $Z_{ij}$ given $\boldsymbol{Y}_j = \boldsymbol{y}_j$ is*

$$E(Z_{ij} = 1 \mid \boldsymbol{y}_j) = \frac{\pi_i \psi(\boldsymbol{y}_j; \boldsymbol{\theta}_i)}{f(\boldsymbol{y}_j; \boldsymbol{\Theta})} \tag{13}$$

(b) *Some specific conditional expectations related to $W_j$ and $U_j$ are*

$$E(W_j \mid \cdots) = (1 - \boldsymbol{\alpha}_i^\top \boldsymbol{\Omega}_i^{-1} \boldsymbol{\alpha}_i)^{1/2}\left(A_{ij} + \frac{\phi(A_{ij})}{\Phi(A_{ij})}\right), \quad (14)$$

$$E(W_j^2 \mid \cdots) = (1 - \boldsymbol{\alpha}_i^\top \boldsymbol{\Omega}_i^{-1} \boldsymbol{\alpha}_i)\left[1 + A_{ij}\left(A_{ij} + \frac{\phi(A_{ij})}{\Phi(A_{ij})}\right)\right], \quad (15)$$

$$E(\tilde{\boldsymbol{U}}_{ij} \mid \cdots) = \boldsymbol{C}_i\big(\boldsymbol{v}_{ij} + \boldsymbol{\lambda}_i(E(W_j \mid \cdots) - c)\big), \quad (16)$$

$$E(W_j \tilde{\boldsymbol{U}}_{ij} \mid \cdots) = \boldsymbol{C}_i\Big\{\boldsymbol{v}_{ij} E(W_j \mid \cdots)$$
$$+ \boldsymbol{\lambda}_i\big[E(W_j^2 \mid \cdots) - cE(W_j \mid \cdots)\big]\Big\}, \quad (17)$$

*and*

$$E(\tilde{\boldsymbol{U}}_{ij}\tilde{\boldsymbol{U}}_{ij}^\top \mid \cdots) = \{\boldsymbol{I}_q + E(\tilde{\boldsymbol{U}}_{ij} \mid \cdots)\boldsymbol{v}_{ij}^\top$$
$$+ \big[E(W_j \tilde{\boldsymbol{U}}_{ij} \mid \cdots) - cE(\tilde{\boldsymbol{U}}_{ij} \mid \cdots)\big]\boldsymbol{\lambda}_i^\top\}\boldsymbol{C}_i, \quad (18)$$

*where* $A_{ij} = (1 - \boldsymbol{\alpha}_i^\top \boldsymbol{\Omega}_i^{-1} \boldsymbol{\alpha}_i)^{-1/2} a_{ij}$.

### 4.2 ML estimation via the ECM algorithm

The EM algorithm has several attractive features such as simplicity of implementation and monotonic convergence properties. However, to compute ML estimates of the MSNFA model, the EM algorithm cannot be directly applied because the M-step is difficult to compute. To go further, we exploit an variant of the EM algorithm, called the ECM algorithm as proposed by Meng and Rubin (1993), which is easy to implement and more broadly applicable than EM. The key feature of ECM is to replace the M-step of EM with a sequence of simpler constrained or conditional maximization (CM) steps. Moreover, it shares all appealing features of EM and can show faster convergence in terms of number of iterations or total computer time.

For notational convenience, let $\boldsymbol{u} = (\boldsymbol{u}_1^\top, \ldots, \boldsymbol{u}_n^\top)^\top$, $\boldsymbol{w} = (w_1, \ldots, w_n)^\top$ and $\boldsymbol{z} = (\boldsymbol{z}_1^\top, \ldots, \boldsymbol{z}_n^\top)^\top$, which are treated as missing data in the EM framework. According to (11), the log-likelihood function for $\boldsymbol{\Theta}$ that can be formed from the complete-data vector $\boldsymbol{y}_c = (\boldsymbol{y}^\top, \boldsymbol{u}^\top, \boldsymbol{w}^\top, \boldsymbol{z}^\top)^\top$, is aside from additive terms not involving the parameters,

$$\ell_c(\Theta; \boldsymbol{y}_c) = \sum_{i=1}^{g}\sum_{j=1}^{n} z_{ij}\Big\{ \log \pi_i - \frac{1}{2}\big[\log|\boldsymbol{D}_i| + \text{tr}(\boldsymbol{D}_i^{-1}\boldsymbol{\Upsilon}_{ij}) \\ + (w_j - c)^2 \boldsymbol{\lambda}_i^\top \boldsymbol{\lambda}_i - 2(\gamma_j - c)\boldsymbol{\lambda}_i^\top \tilde{\boldsymbol{u}}_{ij}\big]\Big\}, \qquad (19)$$

where $\boldsymbol{\Upsilon}_{ij} = (\boldsymbol{y}_j - \boldsymbol{\mu}_i - \tilde{\boldsymbol{B}}_i\tilde{\boldsymbol{u}}_{ij})(\boldsymbol{y}_j - \boldsymbol{\mu}_i - \tilde{\boldsymbol{B}}_i\tilde{\boldsymbol{u}}_{ij})^\top$.

In the E-step of the algorithm, we need to calculate the $Q$-function, denoted by $Q(\Theta; \hat{\Theta}^{(k)})$, which is the conditional expectation of (19) given the observed data $\boldsymbol{y}$ and the current estimates $\hat{\Theta}^{(k)}$. To evaluate the $Q$-function, the necessary conditional expectations include $\hat{z}_{ij}^{(k)} = E(Z_{ij} \mid \boldsymbol{y}_j, \hat{\Theta}^{(k)})$, $\hat{w}_{1ij}^{(k)} = E(W_j \mid Z_{ij} = 1, \boldsymbol{y}_j, \hat{\Theta}^{(k)})$, $\hat{w}_{2ij}^{(k)} = E(W_j^2 \mid Z_{ij} = 1, \boldsymbol{y}_j, \hat{\Theta}^{(k)})$, $\hat{\boldsymbol{\kappa}}_{ij}^{(k)} = E(W_j \tilde{\boldsymbol{U}}_{ij} \mid \boldsymbol{y}_j, \hat{\Theta}^{(k)})$, $\hat{\boldsymbol{\eta}}_{ij}^{(k)} = E(\tilde{\boldsymbol{U}}_{ij} \mid \boldsymbol{y}_j, \hat{\Theta}^{(k)})$ and $\hat{\boldsymbol{\Psi}}_{ij}^{(k)} = E(\tilde{\boldsymbol{U}}_{ij}\tilde{\boldsymbol{U}}_{ij}^\top \mid \boldsymbol{y}_j, \hat{\Theta}^{(k)})$. Therefore, we have

$$Q(\Theta; \hat{\Theta}^{(k)}) = \sum_{i=1}^{g}\sum_{j=1}^{n} \hat{z}_{ij}^{(k)}\Big\{ \log w_j - \frac{1}{2}\big[\log|\boldsymbol{D}_i| + \text{tr}(\boldsymbol{D}_i^{-1}\boldsymbol{\Upsilon}_{ij}^{(k)}) \\ + \hat{h}_{ij}^{(k)} \boldsymbol{\lambda}_i^\top \boldsymbol{\lambda}_i - 2\boldsymbol{\lambda}_i^\top \hat{\boldsymbol{\zeta}}_{ij}^{(k)}\big]\Big\}, \qquad (20)$$

where $\hat{h}_{ij}^{(k)} = \hat{w}_{2ij}^{(k)} - 2c\hat{w}_{1ij}^{(k)} + c^2$, $\hat{\boldsymbol{\zeta}}_{ij}^{(k)} = \hat{\boldsymbol{\kappa}}_{ij}^{(k)} - c\hat{\boldsymbol{\eta}}_{ij}^{(k)}$ and

$$\boldsymbol{\Upsilon}_{ij}^{(k)} = (\boldsymbol{y}_j - \boldsymbol{\mu}_i - \tilde{\boldsymbol{B}}_i\hat{\boldsymbol{\eta}}_{ij}^{(k)})(\boldsymbol{y}_j - \boldsymbol{\mu}_i - \tilde{\boldsymbol{B}}_i\hat{\boldsymbol{\eta}}_{ij}^{(k)})^\top + \tilde{\boldsymbol{B}}_i(\hat{\boldsymbol{\Psi}}_{ij}^{(k)} - \hat{\boldsymbol{\eta}}_{ij}^{(k)}\hat{\boldsymbol{\eta}}_{ij}^{(k)\top})\tilde{\boldsymbol{B}}_i^\top, \quad (21)$$

which involves free parameters $\boldsymbol{\mu}_i$ and $\tilde{\boldsymbol{B}}_i$ for $i = 1, \ldots, g$.

In summary, the implementation of the ECM algorithm proceeds as follows:

**E-step:** Given $\Theta = \hat{\Theta}^{(k)}$, compute $\hat{z}_{ij}^{(k)}, \hat{w}_{1ij}^{(k)}, \hat{w}_{2ij}^{(k)}, \hat{\boldsymbol{\kappa}}_{ij}^{(k)}, \hat{\boldsymbol{\eta}}_{ij}^{(k)}$ and $\hat{\boldsymbol{\Psi}}_{ij}^{(k)}$ by using (13)-(18), for $i = 1, \ldots, g$ and $j = 1, \ldots, n$.

**CM-step 1:** Calculate $\pi_i^{(k+1)} = \hat{n}_i^{(k)}/n$, where $\hat{n}_i^{(k)} = \sum_{j=1}^{n} \hat{z}_{ij}^{(k)}$.

**CM-step 2:** Update $\hat{\boldsymbol{\mu}}_i^{(k)}$ by maximizing (20) over $\boldsymbol{\mu}_i$, which gives

$$\hat{\boldsymbol{\mu}}_i^{(k+1)} = \frac{1}{\hat{n}_i^{(k)}} \sum_{j=1}^{n} \hat{z}_{ij}^{(k)}\Big(\boldsymbol{y}_j - \hat{\tilde{\boldsymbol{B}}}_i^{(k)}\hat{\boldsymbol{\eta}}_{ij}^{(k)}\Big).$$

**CM-step 3:** Fix $\boldsymbol{\mu}_i = \hat{\boldsymbol{\mu}}_i^{(k+1)}$, update $\tilde{\boldsymbol{B}}_i^{(k)}$ by maximizing (20) over $\tilde{\boldsymbol{B}}_i$, which gives

$$\hat{\tilde{\boldsymbol{B}}}_i^{(k+1)} = \sum_{j=1}^n \hat{z}_{ij}^{(k)}\left[(\boldsymbol{y}_j - \hat{\boldsymbol{\mu}}_i^{(k+1)})\hat{\boldsymbol{\eta}}_{ij}^{(k)\top}\right]\left(\sum_{j=1}^n \hat{z}_{ij}^{(k)}\hat{\boldsymbol{\Psi}}_{ij}^{(k)}\right)^{-1}.$$

**CM-step 4:** Fix $\boldsymbol{\mu} = \hat{\boldsymbol{\mu}}_i^{(k+1)}$ and $\tilde{\boldsymbol{B}}_i = \tilde{\boldsymbol{B}}_i^{(k+1)}$, update $\hat{\boldsymbol{D}}_i^{(k)}$ by maximizing (20) over $\boldsymbol{D}_i$, which leads to

$$\hat{\boldsymbol{D}}_i^{(k+1)} = \frac{1}{\hat{n}_i^{(k)}}\mathrm{Diag}\left(\sum_{j=1}^n \hat{z}_{ij}^{(k)}\hat{\boldsymbol{\Upsilon}}_{ij}^{(k)}\right),$$

where $\hat{\boldsymbol{\Upsilon}}_{ij}^{(k)}$ is $\boldsymbol{\Upsilon}_{ij}^{(k)}$ in (21) with $(\boldsymbol{\mu}_i, \tilde{\boldsymbol{B}}_i)$ replaced by $(\hat{\boldsymbol{\mu}}_i^{(k+1)}, \hat{\tilde{\boldsymbol{B}}}_i^{(k+1)})$, respectively.

**CM-step 5:** Update $\hat{\boldsymbol{\lambda}}_i^{(k)}$ by maximizing (20) over $\boldsymbol{\lambda}_i$, which gives

$$\hat{\boldsymbol{\lambda}}_i^{(k+1)} = \frac{\sum_{j=1}^n \hat{z}_{ij}^{(k)}\hat{\boldsymbol{\zeta}}_{ij}^{(k)}}{\sum_{j=1}^n \hat{z}_{ij}^{(k)}\hat{h}_{ij}^{(k)}}.$$

The E- and CM-steps are alternated repeatedly until a suitable convergence rule is satisfied, e.g., the difference in successive values of the log-likelihood is less than a tolerance value. Upon convergence, the ML estimate of $\boldsymbol{\Theta}$ is denoted by $\hat{\boldsymbol{\Theta}} = \{\hat{\pi}_i, \hat{\boldsymbol{\mu}}_i, \hat{\boldsymbol{B}}_i, \hat{\boldsymbol{D}}_i, \hat{\boldsymbol{\lambda}}_i\}_{i=1}^g$, where $\hat{\boldsymbol{B}}_i = \hat{\tilde{\boldsymbol{B}}}_i\hat{\boldsymbol{\Delta}}_i^{1/2}$ and $\hat{\boldsymbol{\Delta}}_i = \boldsymbol{I}_q + (1-c^2)\hat{\boldsymbol{\lambda}}_i\hat{\boldsymbol{\lambda}}_i^\top$. Consequently, the *conditional prediction* of factor scores are estimated by

$$\hat{\boldsymbol{U}}_j = \sum_{i=1}^g \hat{\pi}_i\hat{\boldsymbol{\Delta}}_i^{-1/2}\hat{\boldsymbol{\eta}}_{ij}, \tag{22}$$

where $\hat{\boldsymbol{\eta}}_{ij} = E(\tilde{\boldsymbol{U}}_j \mid Z_{ij} = 1, \boldsymbol{y}_j, \hat{\boldsymbol{\Theta}})$ can be calculated through (16).

### 4.3 Computing standard errors via numerical differentiation

The asymptotic covariance matrix of the ML estimator can be approximated by the inverse of the observed information matrix; see Efron and Hinkley (1978). Specifically, the observed information matrix

$$\boldsymbol{I}(\hat{\boldsymbol{\Theta}}; \boldsymbol{y}) = -\frac{\partial^2 \ell(\boldsymbol{\Theta}; \boldsymbol{y})}{\partial \boldsymbol{\Theta}\partial \boldsymbol{\Theta}^\top}\bigg|_{\boldsymbol{\Theta}=\hat{\boldsymbol{\Theta}}}$$

is a $m \times m$ matrix of second-order partial derivatives of the negative of the log-likelihood function with respect to each parameter, where $m$ is the number of distinct parameters in $\boldsymbol{\Theta}$. For ML theory of large samples, the asymptotic standard errors of $\hat{\boldsymbol{\Theta}}$ can be calculated by taking the square roots of the diagonal elements of $[\boldsymbol{I}(\hat{\boldsymbol{\Theta}};\boldsymbol{y})]^{-1}$.

In the literature, there have been a few strategies recommended for computing $\boldsymbol{I}(\hat{\boldsymbol{\Theta}};\boldsymbol{y})$ efficiently when implementing the EM algorithm; see, for example, Louis (1982) and Meng and Rubin (1991). A problem of these methods is that they require the second-order derivatives of the $Q$-function, which is rather cumbersome to calculate in FA models.

To approximate $\boldsymbol{I}(\hat{\boldsymbol{\Theta}};\boldsymbol{y})$ numerically, Jamshidian (1997) suggested using the central difference. Let $\boldsymbol{s}(\boldsymbol{\Theta};\boldsymbol{y}) = \partial \ell(\boldsymbol{\Theta};\boldsymbol{y})/\partial \boldsymbol{\Theta}$ be the score vector of $\ell(\boldsymbol{\Theta};\boldsymbol{y})$ and $\boldsymbol{s}_c(\boldsymbol{\Theta};\boldsymbol{y}) = \partial \ell_c(\boldsymbol{\Theta};\boldsymbol{y}_c)/\partial \boldsymbol{\Theta}$ be the complete-data score of $\ell_c(\boldsymbol{\Theta};\boldsymbol{y}_c)$. Moreover, it can be verified that $\boldsymbol{s}(\boldsymbol{\Theta};\boldsymbol{y}) = E_{\hat{\boldsymbol{\Theta}}}[\boldsymbol{s}_c(\boldsymbol{\Theta};\boldsymbol{y}_c) \mid \boldsymbol{y}_c]$, see McLachlan and Peel (2000). Explicit expressions for the elements of $\boldsymbol{s}(\boldsymbol{\Theta};\boldsymbol{y})$ are shown below.

$$\boldsymbol{s}_{\pi_r} = \frac{\sum_{j=1}^n \hat{z}_{rj}}{\pi_r} - \frac{\sum_{j=1}^n \hat{z}_{gj}}{\pi_g}, \quad (r=1,\ldots,g-1),$$

$$\boldsymbol{s}_{\boldsymbol{\mu}_i} = \boldsymbol{D}_i^{-1}\left[\sum_{j=1}^n \hat{z}_{ij}(\boldsymbol{y}_j - \boldsymbol{B}\hat{\boldsymbol{\eta}}_{ij}) - \hat{n}_i \boldsymbol{\mu}_i\right],$$

$$\boldsymbol{s}_{\boldsymbol{b}_i} = \mathrm{vec}\left(\boldsymbol{D}_i^{-1}\left\{\sum_{j=1}^n \hat{z}_{ij}(\boldsymbol{y}_j - \boldsymbol{\mu}_i)\hat{\boldsymbol{\eta}}_{ij}^\top - \boldsymbol{B}_i \sum_{j=1}^n \hat{z}_{ij}\hat{\boldsymbol{\Psi}}_{ij}\right\}\right),$$

$$\boldsymbol{s}_{\boldsymbol{d}_i} = \mathrm{diag}\left(-\frac{1}{2}\left(\hat{n}_i \boldsymbol{D}_i^{-1} - \sum_{j=1}^n \hat{z}_{ij}\boldsymbol{D}_i^{-1}\hat{\boldsymbol{\Upsilon}}_{ij}\boldsymbol{D}_i^{-1}\right)\right),$$

and

$$\boldsymbol{s}_{\boldsymbol{\lambda}_i} = \hat{n}_i \frac{(1-c^2)\boldsymbol{\lambda}_i}{1+(1-c^2)\boldsymbol{\lambda}_i^\top \boldsymbol{\lambda}_i} + \sum_{j=1}^n \hat{z}_{ij}\Big[-(1-c^2)\hat{\boldsymbol{\Psi}}_{ij}\boldsymbol{\lambda}_i - \hat{h}_{ij}\boldsymbol{\lambda}_i$$
$$+ \frac{(1-c^2)\boldsymbol{\lambda}_i^\top \hat{\boldsymbol{\zeta}}_{ij}}{\left(1+(1-c^2)\boldsymbol{\lambda}_i^\top \boldsymbol{\lambda}_i\right)^{1/2}}\boldsymbol{\lambda}_i + \left(1+(1-c^2)\boldsymbol{\lambda}_i^\top \boldsymbol{\lambda}_i\right)^{1/2}\hat{\boldsymbol{\zeta}}_{ij}\Big],$$

for $i = 1,\ldots,g$, where $\boldsymbol{b}_i = \mathrm{vec}(\boldsymbol{B}_i)$ and $\boldsymbol{d}_i = \mathrm{diag}(\boldsymbol{D}_i)$.

Let $\boldsymbol{G} = [\boldsymbol{g}_1 \mid \cdots \mid \boldsymbol{g}_m]$ be a $m \times m$ matrix with the $j$th column being

$$\boldsymbol{g}_j = \frac{\boldsymbol{s}(\hat{\boldsymbol{\Theta}} + h_j^* \boldsymbol{e}_j; \boldsymbol{y}) - \boldsymbol{s}(\hat{\boldsymbol{\Theta}} - h_j^* \boldsymbol{e}_j; \boldsymbol{y})}{2h_j^*}, \quad j = 1, \cdots, m,$$

where $\boldsymbol{e}_j$ is a unit vector corresponding to the $j$th element. The values of $h_j^*$ are small numbers chosen based on the scale of problem. In later data analysis, we will use $h_j = \max(\eta, \eta|\hat{\boldsymbol{\Theta}}_j|)$ with $\hat{\boldsymbol{\Theta}}_j$ denoting denoting the $j$th of element of $\hat{\boldsymbol{\Theta}}$, where values such as $\eta = 10^{-4}$ should be small enough to approximate and large enough to avoid the roundoff error. Since $\boldsymbol{G}$ may not be symmetric, it is suggested using

$$\tilde{\boldsymbol{I}}(\hat{\boldsymbol{\Theta}}; \boldsymbol{y}) = -\frac{\boldsymbol{G} + \boldsymbol{G}^\top}{2}$$

to approximate $\boldsymbol{I}(\hat{\boldsymbol{\Theta}}; \boldsymbol{y})$.

# 5 Strategies for implementation

## 5.1 Initialization

As described in Section 4, the MSNFA parameters are estimated through the ECM algorithm. However, the EM-type algorithm has an intrinsic limitation that there is no guarantee of convergence to the global optimum (Wu, 1983). For modeling multi-model distributions, the iterations may converges to a local maximum or to a saddle point. Sometimes, the quality of the final solution depends heavily on starting values. To cope with such potential problems, we recommend a simple way of obtaining suitable initial values for the ECM algorithm below.

1. Perform the $k$-means algorithm initialized with a random seed. Then, initialize the zero-one membership indicator $\hat{\boldsymbol{z}}_j^{(0)} = \{\hat{z}_{ij}^{(0)}\}_{i=1}^g$ according to the $k$-means clustering result. The initial values for the mixing proportions and component locations are then given by

$$\hat{\pi}_i^{(0)} = \frac{\sum_{j=1}^n \hat{z}_{ij}^{(0)}}{n} \quad \text{and} \quad \hat{\boldsymbol{\mu}}_i^{(0)} = \frac{\sum_{j=1}^n \hat{z}_{ij}^{(0)} \boldsymbol{y}_j}{\sum_{j=1}^n \hat{z}_{ij}^{(0)}}.$$

**2.** Subtract each observation from its initial cluster means. Then, do a FA fit to these $k$ "centering samples" via the ML estimation (default) or the PCA method. The resulting estimates of factor loading and error covariance matrices are taken as initial values, namely $\hat{\boldsymbol{B}}_i^{(0)}$ and $\hat{\boldsymbol{D}}_i^{(0)}$ for $i = 1, \ldots, g$. Next, compute the corresponding factor scores of each cluster via the *conditional prediction* method such as (22). The initial values for the skewness parameters $\hat{\boldsymbol{\lambda}}_i^{(0)}$ are obtained by fitting the rMSN distribution to the $k$ samples of factor scores via the R package EMMIX-skew (Wang, 2009).

The above procedure provides a quick and convenient strategy to initialize the parameters. Once the EM algorithm has converged, we can determine the cluster membership according to the maximum *a posteriori* (MAP) classification rule. That is, each observation $\boldsymbol{y}_j$ is assigned to the component with the highest posterior probability.

The ECM procedure can get stuck in one of the many local maxima of the likelihood function (Meng and Rubin, 1993). To overcome such a flaw, it is recommended to initialize the algorithm with various choices of starting values for searching for all local maxima (McLachlan and Krishnan, 2008). This can be done by specifying a variety of other starting points such as *random starts* (McLachlan and Peel, 2000) or model-based hierarchical clustering methods (Fraley, 1998). The global optimum $\hat{\boldsymbol{\Theta}}$ can be the one which has the highest log-likelihood value.

### 5.2 Model selection

A number of information criteria have been proposed to facilitate identifying an appropriate model. The most frequently employed index is the Bayesian Information Criterion (BIC; Schwarz, 1978)

$$\text{BIC} = \ell_{\max} - \frac{m}{2} \log n,$$

where $m$ is the number of free parameters and $\ell_{\max}$ is the maximized log-likelihood value. Empirical evidence has shown that BIC is useful in choosing the true number of classes of a given mixture model and an ideal number of latent factors, e.g., McNicholas and Murphy (2008), Baek et al. (2010) and Baek and McLachlan (2011).

As outlined by Biernacki et al. (2000), an alternative promising measure for estimating the proper number of clusters is based on the integrated completed likelihood (ICL), defined as

$$\text{ICL} = \text{BIC} - \text{ENT}(\hat{\boldsymbol{z}}),$$

where $\text{ENT}(\hat{\boldsymbol{z}}) = -\sum_{i=1}^{g}\sum_{j=1}^{n}\hat{z}_{ij}\log\hat{z}_{ij}$ is the entropy of the classification matrix with the $(i,j)th$ equal to $\hat{z}_{ij}$ with $\hat{z}_{ij}$ being the posterior probability of $\boldsymbol{y}_j$ classified to class $i$. Simply speaking, ICL is equal to BIC penalized by subtracting the estimated mean entropy, which is used to measure the overlap of clusters. It penalizes complex models more severely than BIC and thus favors models with fewer latent classes, providing a better estimate of the number of well-separated clusters.

When ICL leads too many factors being fitted in the mixtures, we have also considered the approximate weight of evidence (AWE; Banfield and Raftery, 1993), given by

$$\text{AWE} = \text{ICL} - m(3/2 + \log n),$$

which places a higher penalty than ICL on more complex model due to the the extra constant term $m(3/2 + \log n)$. In general, larger BIC, ICL or AWE values indicate a better fitted model. We note by passing that there is no clear consensus regarding which criterion is better to use. This depends on the problem at hand and usually a combined use would be of help to screen reasonable candidate models.

### 5.3 Convergence assessment

To monitor the convergence by using the likelihood increasing property of the ECM algorithm, we recommend employing the simplest stoping rule $\ell(\boldsymbol{\Theta}^{(k)}|\boldsymbol{y}) - \ell(\boldsymbol{\Theta}^{(k-1)}|\boldsymbol{y}) < \epsilon$, where $\epsilon$ is a user-specified tolerance. Another recommendation is

to adopt the Aitken's acceleration criterion (McLachlan and Krishnan, 2008) which estimates the asymptotic maximum of the likelihood and allows to detect an early convergence. In our analysis, the algorithm is terminated if the maximum number of iterations $k_{\max}$ =5,000 is reached or when the difference between two successive log-likelihood values is less than $\epsilon = 10^{-6}$.

### 5.4 Performance evaluation

To assess the model-based classification accuracy, we compute the correct classification rate (CCR) and the adjusted Rand index (ARI) as proposed by Hubert and Arabie (1985). The CCR is calculated by considering all permutations of the class labels and the one with the lowest misclassification error was treated as the final class membership assignment. As a measure of class agreement, the ARI account for the fact that a random classification may correctly classify some cases. Note that the ARI has expected values of 0 under random classification and 1 for perfect classification. For both CCR and ARI, larger values indicate better classification results.

## 6 Application

### 6.1 The AIS dataset

As a simple illustration, we consider the Australian Institute of Sport (AIS) data (Cook and Weisberg, 1994; Azzalini and Capitanio, 1999) containing $p = 11$ physical and hematological attributes measured on $n = 202$ athletes (100 female and 102 male). The dataset is publicly available from the R package `sn` (Azzalini, 2011). A detailed account of these attributes along with their sample skewness and kurtosis (a split by gender) are separately summarized in Table 1. Among these attributes, most of them are moderately to strongly skewed and are highly leptokurtotic for both genders.

**Table 1 about here**

Before proceeding the fitting, the data has been standardized such that each variable has zero mean and unit standard deviation to avoid some variables having a greater impact due to their different scales. To explore the unsupervised learning, both MFA and MSNFA models were fitted to the data for $g=1$–$5$ and $q=1$–$5$. The values of BIC as well as ICL and AWE under each scenario was computed and the best selected model was MFA ($g=4$ and $q=4$) with the highest BIC value of –1131.2, followed by MSNFA ($g=3$ and $q=4$) with a BIC value of –1150.1. Figure 2 shows the heat map of BIC values for each pair $(g,q)$. The number of free parameters is 243 for MFA and 194 for MSNFA, respectively. Observing these results, we prefer using the MSNFA approach for this dataset as it uses fewer free parameters and leads to fewer components $g=3$ (regarded as classified groups), which is closer to the true number of clusters, that is $g=2$. Tables 2 and 3 list a cross-tabulation of the MAP classification for the best two models versus the true memberships. Clearly, MSNFA has a better classification performance than does MFA in separating two intrinsic groups. Note that the comparison results are similar for ICL and AWE.

**Figure 1 about here**

Since the group labels (athlete's gender) are provided in advance, we implement the 2-component MFA and MSNFA with different levels $q=1-6$. Here the choice of maximum $q=6$ satisfies the restriction $(p-q)^2 \leq (p+q)$ as suggested by Eq. (3) of Fokoué and Titterington (2003). From the BIC curves shown in Figure 2(a), the fitting performance between the two models are comparable. The best model is taken as MSNFA ($q=4$) which attains the largest BIC (–1172.6). A summary of ML fitting results is given in Table 4. Some of the estimated skewness parameters are moderately significant, revealing that the joint distribution of this dataset, to some extent, departs from normality. Notice that both MFA and MSNFA yield equally better classification accuracies (details are not shown). The reason is partly due to the fact that the effect of skewness does not lead to a high impact on classification in this example.

**Tables 2 and 3 about here**

We have also given coordinate projected plots for a subset of AIS data under the best selected MSNFA. These plots are commonly employed to exhibit a graphical display of two-dimensional presentation. Figure 2(b) displays a coordinate projected classification plot of the data for variables `bmi` and `Bfat`. Figure 2(c) depicts the uncertainty plot, obtained by subtracting the probability of the most likely group for each observation from one. Figure 2(d) shows the corresponding two-group classification plot, which matches the true group labels with six misclassified units.

**Table 4 about here**

**Figure 2 about here**

### 6.2 The WDBC dataset

Breast cancer is a major cause of death among women. Early detection of breast cancer through classification can avoid unnecessary surgery. As another illustration, we applied our method to the Wisconsin Diagnostic Breast Cancer (WDBC) data, which are available from the UCI Machine Learning data repository (Frank and Asuncion, 2010). These data consist of $n = 569$ instances with a total of 32 different attributes. The first two attributes correspond to the ID number which will not be used and the diagnosis status, of which 357 have the diagnosis benign and 212 have the diagnose malignant. The rest $p = 30$ attributes are ten real-valued measurements (Radius, Texture, Perimeter, Area, Smoothness, Compactness, Concavity, Concave points, Symmetry and Fractal dimension) computed from a digitized mammography image of a fine needle aspirates (FNA) of breast tissue, together with their associated mean, standard error and the mean of the three largest ('worst') values, respectively.

Table 5 about here

Since there are two known classes, we implemented two-component MFA and MSNFA models with $q$ ranging from 1–10. To fit the models via the ML method, the ECM algorithm developed in Section 4.2 was employed under twenty different initializations for the parameters. The resulting ML solutions, including the maximized log-likelihood values, the number of parameters together with the BIC, ICL and AWE values are listed in Table 5. To compare the classification accuracy, we also computed the ARI and CCR for each $q$. As can be seen, the best fitted model is MSNFA with $q = 9$, no matter which selection criterion was used. In addition, the MSNFA yields the best ARI (0.762) and CCR (0.937) when $q = 7$. The result confirms that the MSNFA is more appropriate for this dataset, providing more accurate classification accuracies when the data exhibit a departure from normality.

# 7 Conclusion

We introduce the MSNFA model obtained from the classical MFA model by replacing the normal latent factors with the rMSN distribution for each component. This family of mixture analyzers has emerged as an attractive tool since it can account for groups of data exhibiting patterns of asymmetry and multimodality which are commonly seen in high-dimensional data. For estimating parameters, an analytically simple ECM algorithm is developed under a four-level hierarchical framework. Some computational strategies related to the specification of starting values, convergence assessment and provision of standard errors are provided. Two main identification problems regarding invariant likelihood caused by factor indeterminacy and label switching are also discussed. We should mention that both of which do not affect the clustering results. Numerical results through the information-based model selection and classification accuracy indicate the effectiveness and superiority of the proposed method when compared with ordinary MFA ones.

There are a number of possible extensions of the current work. While the proposed MSNFA has shown its great flexibility in modeling asymmetric features among heterogeneous data, its robustness against outliers could still be unduly influenced by heavy-tailed observations. Mixtures of factor analyzers based on more general forms such as the skew $t$-distribution and its variants (Azzalini and Capitaino, 2003; Sahu et al., 2003; Pyne et al., 2009) would be of interest for future research. Another worthwhile task is to develop workable Markov chain Monte Carlo algorithms (Hastings, 1970; Tanner and Wong, 1987; Diebolt and Robert, 1994; Escobar and West, 1995) for drawing much richer inferences under a Bayesian paradigm. Although the proposed ECM procedure is quite easy to implement, its convergence can be painfully slow in certain situations. Therefore, it is also of interest to pursue some modified algorithms toward fast convergence, see, for instance, Zhao and Yu (2008) and Wang and Lin (2013).

Table 1: An overview of 11 attributes of the AIS data

| Variable | Description | Female | | Male | |
|---|---|---|---|---|---|
| | | Skewness | Kurtosis | Skewness | Kurtosis |
| rcc | red cell count | 0.69 | 3.3 | 0.92 | 7.73 |
| wcc | white cell count | 0.75 | 4 | 0.86 | 4.58 |
| Hc | Hematocrit | 0.26 | 2.34 | 1.49 | 10.37 |
| Hg | Hemoglobin | 0.09 | 2.18 | 0.97 | 5.31 |
| Fe | plasma ferritin concentration | 1.35 | 5.57 | 0.88 | 3.13 |
| bmi | body mass index | 0.69 | 4.18 | 1.41 | 5.99 |
| ssf | sum of skin folds | 0.78 | 3.64 | 1.39 | 4.79 |
| Bfat | body fat percentage | 0.35 | 2.91 | 1.53 | 5.08 |
| lbm | lean body mass | −0.31 | 3.45 | 0.27 | 3.62 |
| Ht | height (cm) | −0.56 | 4.2 | 0.07 | 3 |
| Wt | weight (Kg) | −0.17 | 3.13 | 0.39 | 3.41 |

Table 2: A classification table for the MFA ($g = 4$; $q = 3$) on the AIS data

| MFA | Cluster | | | |
|---|---|---|---|---|
| | 1 | 2 | 3 | 4 |
| female | 57 | 0 | 39 | 4 |
| male | 0 | 80 | 5 | 17 |

Table 3: A classification table for the MSNFA ($g = 3$; $q = 3$) on the AIS data

| MSNFA | Cluster | | |
|---|---|---|---|
| | 1 | 2 | 3 |
| female | 58 | 0 | 42 |
| male | 0 | 90 | 12 |

Table 4: Summary ML results together with the associated standard errors in parentheses for the best chosen model.

| Variable | class 1 | | | | | | class 2 | | | | | |
|---|---|---|---|---|---|---|---|---|---|---|---|---|
| | $\boldsymbol{\mu}_1$ | | $\boldsymbol{B}_1$ | | | $\boldsymbol{d}_1$ | $\boldsymbol{\mu}_2$ | | $\boldsymbol{B}_2$ | | | $\boldsymbol{d}_2$ |
| rcc | -0.68 | 0.6 | -0.05 | 0.04 | -0.03 | 0.13 | 0.68 | -0.02 | 0.58 | 0.29 | 0.01 | 0.13 |
| | (0.08) | (0.06) | (0.08) | (0.16) | (0.19) | (0.02) | (0.07) | (0.06) | (0.2) | (0.12) | (0.14) | (0.02) |
| wcc | -0.09 | 0.2 | 0.21 | -0.03 | 0.06 | 0.79 | 0.13 | -0.03 | -0.04 | 0.29 | -0.09 | 0.97 |
| | (0.11) | (0.11) | (0.11) | (0.1) | (0.11) | (0.14) | (0.11) | (0.11) | (0.11) | (0.15) | (0.1) | (0.12) |
| Hc | -0.7 | 0.7 | -0.1 | 0.07 | -0.01 | 0.00 | 0.7 | 0.00 | 0.6 | 0.31 | 0.04 | 0.02 |
| | (0.07) | (0.06) | (0.09) | (0.17) | (0.19) | (0.01) | (0.07) | (0.06) | (0.23) | (0.13) | (0.18) | (0.01) |
| Hg | -0.73 | 0.6 | -0.07 | 0.03 | 0.07 | 0.07 | 0.73 | -0.01 | 0.54 | 0.28 | 0.13 | 0.07 |
| | (0.07) | (0.08) | (0.1) | (0.15) | (0.18) | (0.01) | (0.07) | (0.07) | (0.21) | (0.1) | (0.16) | (0.01) |
| Fe | -0.41 | -0.05 | 0.03 | -0.13 | -0.02 | 0.38 | 0.46 | -0.08 | -0.12 | 0.15 | 0.27 | 1.11 |
| | (0.12) | (0.13) | (0.14) | (0.13) | (0.11) | (0.17) | (0.06) | (0.07) | (0.09) | (0.07) | (0.07) | (0.07) |
| bmi | -0.42 | -0.01 | 0.53 | 0.16 | 0.61 | 0.00 | 0.49 | 0.46 | -0.11 | 0.49 | 0.62 | 0.00 |
| | (0.09) | (0.23) | (0.27) | (0.09) | (0.17) | (0.00) | (0.06) | (0.17) | (0.04) | (0.02) | (0.03) | (0.00) |
| ssf | 0.48 | -0.18 | 0.98 | 0.24 | 0.06 | 0.05 | -0.4 | 0.05 | -0.22 | 0.47 | 0.07 | 0.02 |
| | (0.05) | (0.04) | (0.14) | (0.07) | (0.07) | (0.00) | (0.1) | (0.23) | (0.09) | (0.14) | (0.17) | (0.01) |
| Bfat | 0.63 | -0.13 | 0.87 | 0.24 | 0.02 | 0.01 | -0.56 | 0.00 | -0.2 | 0.44 | 0.06 | 0.00 |
| | (0.05) | (0.04) | (0.14) | (0.06) | (0.06) | (0.00) | (0.09) | (0.2) | (0.08) | (0.12) | (0.15) | (0.00) |
| lbm | -0.8 | 0.00 | 0.08 | 0.39 | 0.3 | 0.00 | 0.81 | 0.66 | -0.06 | 0.16 | 0.08 | 0.00 |
| | (0.08) | (0.07) | (0.06) | (0.05) | (0.23) | (0.00) | (0.00) | (0.00) | (0.00) | (0.00) | (0.00) | (0.00) |
| Ht | -0.59 | -0.07 | 0.17 | 0.82 | -0.05 | 0.00 | 0.57 | 0.57 | -0.14 | 0.04 | -0.5 | 0.00 |
| | (0.09) | (0.16) | (0.14) | (0.08) | (0.22) | (0.00) | (0.66) | (0.14) | (0.06) | (0.1) | (0.1) | (0.00) |
| Wt | -0.61 | -0.04 | 0.42 | 0.51 | 0.34 | 0.00 | 0.65 | 0.69 | -0.15 | 0.36 | 0.12 | 0.00 |
| | (0.09) | (0.08) | (0.11) | (0.06) | (0.25) | (0.00) | (0.03) | (0.07) | (0.03) | (0.04) | (0.05) | (0.00) |
| | w | | $\boldsymbol{\lambda}_1$ | | | | | | $\boldsymbol{\lambda}_2$ | | | |
| | 0.5 | -0.01 | 2.63 | -0.26 | 0.39 | | | -1.19 | -7.08 | 10.26 | 2.03 | |
| | (0.04) | (0.24) | (4.88) | (2.62) | (1.63) | | | (0.45) | (0.61) | (0.59) | (0.3) | |

Table 5: Comparison of MFA and MSNFA fitting results and implied clustering versus the true membership of WDBC data

| Model | $q$ | $\ell_{\max}$ | $m$ | BIC | ICL | AWE | ARI | CCR |
|---|---|---|---|---|---|---|---|---|
| MFA | 1 | 9624.8 | 181 | 9050.7 | 9041.9 | 8196.2 | 0.520 | 0.861 |
| | 2 | 12362.7 | 239 | 11604.6 | 11596.6 | 10480.0 | 0.396 | 0.817 |
| | 3 | 13962.5 | 295 | 13026.8 | 13021.7 | 11643.4 | 0.359 | 0.803 |
| | 4 | 15616.8 | 349 | 14509.8 | 14506.8 | 12876.3 | 0.658 | 0.907 |
| | 5 | 15726.5 | 401 | 14454.6 | 14448.7 | 12575.3 | 0.595 | 0.888 |
| | 6 | 16691.4 | 451 | 15260.8 | 15256.7 | 13149.6 | 0.630 | 0.898 |
| | 7 | 17017.2 | 499 | 15434.4 | 15431.0 | 13099.7 | 0.670 | 0.910 |
| | 8 | 17248.6 | 545 | 15519.9 | 15515.3 | 12969.1 | 0.595 | 0.888 |
| | 9 | 18467.3 | 589 | 16599.0 | 16595.4 | 13843.6 | 0.700 | 0.919 |
| | 10 | 17692.3 | 631 | 15690.8 | 15685.0 | 12737.0 | 0.624 | 0.896 |
| MSNFA | 1 | 9632.8 | 183 | 9052.4 | 9043.11 | 8188.1 | 0.515 | 0.859 |
| | 2 | 12441.3 | 243 | 11670.5 | 11662.6 | 10527.3 | 0.373 | 0.808 |
| | 3 | 14117.8 | 301 | 13163.1 | 13158.6 | 11752.3 | 0.397 | 0.817 |
| | 4 | 15700.5 | 357 | 14568.1 | 14563.8 | 12895.9 | 0.658 | 0.907 |
| | 5 | 15830.1 | 411 | 14526.5 | 14521.3 | 12601.1 | 0.618 | 0.895 |
| | 6 | 16933.3 | 463 | 15464.7 | 15459.3 | 13296.2 | 0.718 | 0.924 |
| | 7 | 17486.8 | 513 | 15859.6 | 15856.0 | 13459.2 | **0.762** | **0.937** |
| | 8 | 17572.5 | 561 | 15793.0 | 15789.5 | 13168.6 | 0.681 | 0.914 |
| | 9 | **18598.8** | 607 | **16673.5** | **16670.2** | **13834.3** | 0.712 | 0.923 |
| | 10 | 18000.9 | 651 | 15936.0 | 15931.4 | 12890.0 | 0.700 | 0.919 |

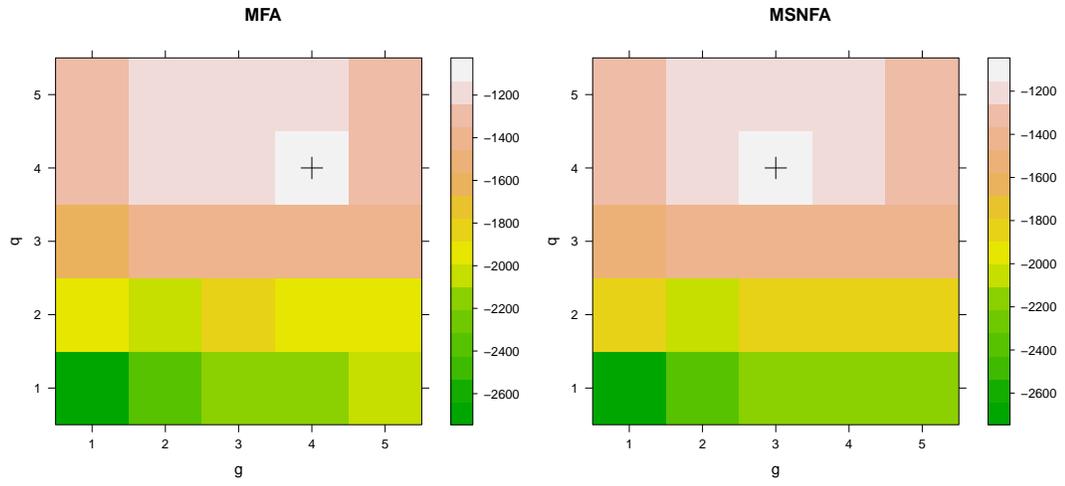

Figure 1: A heat map representation of BIC values of MFA (left panel) and MSNFA (right panel) over each combination of $(g, q)$ for the AIS data

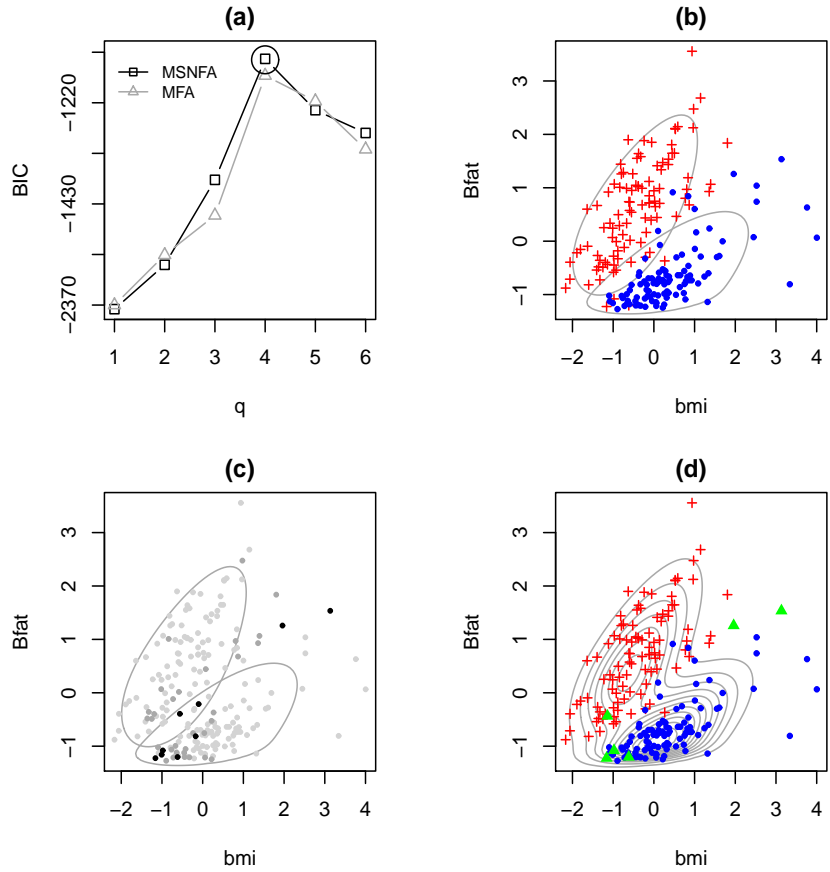

Figure 2: (a) A comparison of BIC values up to 6 factors for the AIS data. The optimal model is taken to be the highest BIC from the fitted models. (b) A projection plot with different symbols indicating the classification corresponding to MSNFA ($q = 4$) (the best model) as determined. The non-elliptically contoured curves are drawn corresponding to their component covariance matrices. (c) The uncertainty plot of a classification. (d) A projection plot showing errors in the classification. Full green symbols indicate incorrectly classified observations.